\documentclass[aps,prl,twocolumn,showpacs,superscriptaddress,groupedaddress]{revtex4}  % for review and submission
\usepackage{graphicx}  % needed for figures
\usepackage{dcolumn}   % needed for some tables
\usepackage{bm}        % for math
\usepackage{amssymb}   % for math
\usepackage{color}
\usepackage{pdfpages}

\begin{document}

\title{Zeeman Splitting of Photonic Angular Momentum States in Gyromagnetic Cylinder}

\author{Jin Wang} \thanks{These authors contributed equally to this work.}\affiliation{Department of Physics, Southeast University, Nanjing 211189, China}\affiliation{Department of Mechanical Engineering, Massachusetts Institute of Technology, Cambridge, MA 02139, USA}
\author{Kin Hung Fung} \thanks{These authors contributed equally to this work.}\affiliation{Department of Mechanical Engineering, Massachusetts Institute of Technology, Cambridge, MA 02139, USA}
\author{Hui Yuan Dong} \affiliation{Department of Physics, Southeast University, Nanjing 211189, China}\affiliation{School of Science, Nanjing University of Posts and Telecommunications, Nanjing 210003, China}
\author{Nicholas X. Fang} \email{nicfang@mit.edu}\affiliation{Department of Mechanical Engineering, Massachusetts Institute of Technology, Cambridge, MA 02139, USA}
\date{\today}

\begin{abstract}

We show that under the presence of a static magnetic field the photon eigen-frequencies of a circular gyromagnetic cylinder experience a splitting that is proportional to the angular momentum density of light at the cylinder surface. Such a splitting of the photonic states is similar to the Zeeman splitting of electronic states in atoms. This leads to some unusual decoupling properties of these non-degenerate photonic angular momentum states, which are demonstrated through numerical simulations.

\end{abstract}

\maketitle
Recently, the effect of static magnetic field on photonic states has attracted a lot of attention due to the discovery of protected photonic chiral edge states in gyromagnetic photonic crystals \cite{haldane,haldane2,zheng}, which are analogue to the topologically protected edge states in electronic systems \cite{Hasan}. However, analytical studies on simple photonic states are still rare, compared to the extensive studies of electronic states. One of the most fundamental examples for electronic states is the famous Zeeman effect \cite{Zeeman}. In analogy with the Zeeman splitting of electronic states in atom, here we study a similar basic effect on photonic states in a gyromagnetic cylinder. We also introduce some unusual wave decoupling phenomena which are consequences of such a splitting.

Since Zeeman splitting of electronic states is associated with the broken degeneracies of electron states with different angular momenta, we expect to see a relation between the Zeeman-like effect in our photonic system and the angular momentum of light. Angular momentum of light \cite{LightAM}, which could be used for storage of quantum information \cite{QuanInfo}, has drawn serious interest in recent years. Therefore, it is our motivation in this Letter to calculate the frequency splitting of photonic angular momentum states and derive a formula which relates the angular momentum of light to the frequency splitting.

\begin{figure}[htbp]
\centering
\includegraphics[width=2.4in]{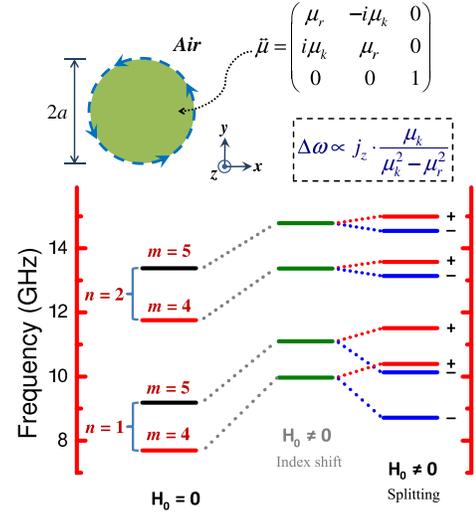}
\caption{\small (color online) Zeeman splitting of photonic states in the presence of static magnetic field. The top figure shows a gyromagnetic cylinder surrounded by air. Formula in the dashed box displays the relation between frequency shift and angular momentum density of light ($j_z$). The first column (black lines) shows the original degenerate states in the gyromagnetic cylinder in the absence of magnetic field $H_{0}=0$ (i.e., with isotropic permeability $\mu_{r}$). The second column (green lines) shows the shifts due to the change of effective permeability from $\mu_{r}$ to $\mu'_{r}$ (indicated as ``index shift'') when a static magnetic field of $H_{0}=800$~Oe is considered. The third column (red and blue lines) shows the final Zeeman splitting with the adjacent signs indicating the sign of $m$. Only the states with $n=1,2$ and $|m|=4,5$ are shown.}
\end{figure}

We consider the splitting of photonic states in gyromagnetic cylinder as shown in Fig. 1. A gyromagnetic material can have a permeability depending on the applied static magnetic field. The magnetic permeability tensor can be written as
\begin{equation}
\bar{\mu}_{c}=
\left(
\begin{tabular}{ccc}
$\mu_{r}$ & $-i\mu_{k}$ & $0$\\
$i\mu_{k}$ & $\mu_{r}$ & $0$\\
$0$ & $0$ & $1$\\
\end{tabular}
\right).
\end{equation}
For transverse electric (TE) polarization with electric field along the $z$ direction, the frequencies ($\omega$) of the photonic states, can be calculated from the roots of the Mie resonance condition \cite{Effective},
\begin{eqnarray}
&\sqrt{\frac{\epsilon_{d}}{\mu'_{r}}}J'_{m}(k'a)H_{m}^{(1)}(k_{0}a)-J_{m}(k'a)H_{m}^{(1)\prime}(k_{0}a) \nonumber\\
&-\frac{m\mu_{k}}{(\mu_{r}^{2}-\mu_{k}^{2})k_{0}a}J_{m}(k'a)H_{m}^{(1)}(k_{0}a)=0, \label{disp}
\end{eqnarray}
where $m$ is the azimuthal quantum number, $J_m$ is the $m$th-order Bessel function, $H_m^{(1)}$ is the $m$th-order Hankel function of the first kind, $k'=k_{0}\sqrt{\mu'_{r}\epsilon_{d}}$, $k_{0}=\omega/c$ is free-space wave number, and $\mu'_{r}=(\mu_{r}^{2}-\mu_{k}^2)/\mu_{r}$. It should be noted that Eq. (\ref{disp}) does not have real root frequencies because of radiation loss but we will only consider the states with relatively low radiation loss (i.e., frequencies with small imaginary parts). By comparing Eq. (\ref{disp}) with the Mie condition for a cylinder with isotropic permeability $\mu_{r}$ [i.e., Eq. (\ref{disp}) with $\mu_{k}=0$], we can interpret the effect of the static magnetic field as two steps: (i) a shift associated with an index change (from permeability $\mu_{r}$ to $\mu'_{r}$) and (ii) a splitting of frequencies.

As an example for numerical demonstration, we consider yttrium-iron-garnet (YIG), which is a type of commercially available gyromagnetic materials, as the material of the cylinder supporting photonic states at microwave frequencies. For an applied static magnetic field $H_{0}$ in the $z$ direction, we have $\mu_{r}=1+\omega_{m}\omega_{h}/(\omega_{h}^{2}-\omega^{2})$ and $\mu_{k}=\omega_{m}\omega/(\omega_{h}^{2}-\omega^{2})$ \cite{MagnetTheory}, where $\omega_{h}=\gamma H_{0}$ is the precession frequency, $\gamma$ is the gyromagnetic ratio, $\omega_{m}=4\pi\gamma M_{s}$, $4\pi M_{s}$ is the saturation magnetization. Using parameters provided in a previous experimental study ($\epsilon_d=15.26$, $H_{0}=800$~Oe, and $4\pi M_{s}=1884~$G) \cite{ExpOneWay}, we plot the frequency splitting diagram in Fig. 1 for a cylinder of radius $a=1~$cm by finding the frequency roots of Eq. (\ref{disp}). Here, in addition to the azimuthal quantum number $m$, we denotes $n$ as the quantum number in the radial direction and ($n$, $m$) as a specific photon state. To have a clear picture, we first focus on $|m|=4,5$ in the lowest ($n=1$) and the second lowest ($n=2$) radially quantized levels. In the absence of the static magnetic field (first column of Fig.~1), the positive-$m$ and negative-$m$ states are degenerate. However, when the static magnetic field is present, the original degenerate states shift up to higher frequencies (second column of Fig.~1) and split into two counter rotating states ($n$, $m$) and ($n$, $-m$), as indicated respectively by the red lines and blue lines in the third column of Fig.~1. Physically, the effect of the static magnetic field can be understood as a broken time reversal symmetry (and reciprocity) so that photonic states are no longer degenerate. We will show in a later part of this Letter that such a splitting is proportional to the angular momentum of light.

\begin{figure}[htbp]
\centering
\includegraphics[width=2.8in]{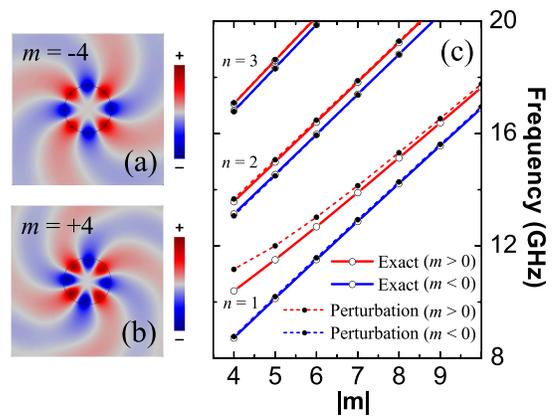}
\caption {\small (color online) (a) and (b) Electric field patterns excited by a line source in the presence of external magnetic field ($H_{0}=800~$Oe). (a) Only $m=-4$ mode is excited at its resonant frequency $f=8.72~$GHz. (b) Same as (a), but for $m=+4$ at its resonant frequency $f=10.40~$GHz.
(c) Resonant frequency versus azimuthal momentum number $m$. The solid lines with open circles show exact results obtained from Mie resonant conditions [Eq. (\ref{disp})].  The dashed lines with solid circles indicate approximated results obtained from the perturbation theory [Eq. (\ref{ClosedForm})]. Blue lines and red lines indicate the negative $m$ modes and the positive $m$ modes, respectively. Low-quality states with $|m|<4$ are not shown.}
\end{figure}

Using a finite element solver (COMSOL Multiphysics), we examine the electric field profile of the non-degenerate rotational states after splitting. Fig. 2(a) and (b) show the wave profiles excited by a line current source lying along the $z$-direction on the surface of gyromagnetic cylinder at frequencies $f=8.72~$GHz and $f=10.40~$GHz, respectively. In Fig. 2(a), we see a clockwise rotating field pattern with four complete oscillations along the azimuthal directions, which shows that only the $(1, -4)$ state can be excited at $f=8.72~$GHz. The field pattern at $f=10.40~$GHz [Fig. 2(b)] is similar to that in Fig. 2(a) except that it is for $(1, 4)$ state with its electric field rotating in anti-clockwise direction.  These results confirm the Zeeman splitting diagram in Fig. 1.

For a complete picture of the splitting of photonic states, we plot in Fig.~2(c) the exact solutions [roots of Eq. (\ref{disp})] for the resonant frequency as a function of the azimuthal momentum number $m$. It is found that the resonant frequencies show pronounced differences between positive and negative $m$ states, stemmed from the effect of the static magnetic field. For $n=1$, the most obvious frequency splitting can be observed and the splitting gap width decreases gradually as $m$ increases. We also observe that the splitting is relatively weak for larger $n$ and, for each fixed $n$, the splitting is almost independent of $m$ when $m$ is large.

To understand the frequency splitting and its relation to the angular momentum of light, we employ a Hamiltonian eigenvalue approach for electromagnetic waves \cite{Hamilton1,Hamilton2}. In this approach, we construct the following eigenvalue problem:
\begin{equation}
\Omega |\psi\rangle=\omega |\psi\rangle \label{Hamit}
\end{equation}
where
\begin{equation}
\Omega\doteq
\left(
\begin{tabular}{ccc}
$0$ & $\frac{i}{\epsilon_{0}}\bar{\epsilon}^{-1}\nabla\times$ \\
$-\frac{i}{\mu_{0}}\bar{\mu}^{-1}\nabla\times$ & $0$\\
\end{tabular}
\right),
|\psi\rangle\doteq
\left(
\begin{tabular}{ccc}
$\vec{E}$ \\
$\vec{H}$
\end{tabular}
\right),
\end{equation}
$\bar{\epsilon}(\vec{r})$ and $\bar{\mu}(\vec{r})$ are, respectively, the position dependent relative permittivity and permeability tensor, $\vec{E}(\vec{r})$ and $\vec{H}(\vec{r})$ are, respectively, the dynamic electric and magnetic fields. Here, we take the approximation that the dispersive material properties is weak in the frequency range where $\omega$ is much larger than $\omega_{h}$. For TE polarization, we can simplify Eq. (\ref{Hamit}) as
\begin{equation}
\left(
\Omega_{0}+\frac{\mu_{k}}{\mu_{k}^{2}-\mu_{r}^{2}}\Delta\Omega
\right)|\psi\rangle=\omega|\psi\rangle
\end{equation}
where $|\psi\rangle\doteq (E_{z},H_{r}, H_{\phi})^{T}$,
\begin{equation}
\Omega_{0}\doteq i\left(
\begin{tabular}{ccc}
$0$ & $-\frac{1}{\epsilon_{0}\epsilon(r)}\frac{1}{r}\frac{\partial}{\partial\phi}$ & $\frac{1}{\epsilon_{0}\epsilon(r)}(\frac{1}{r}+\frac{\partial}{\partial r})$\\
$-\frac{1}{\mu_{0}\mu(r)}\frac{1}{r}\frac{\partial}{\partial\phi}$ & $0$ & $0$\\
$\frac{1}{\mu_{0}\mu(r)}\frac{\partial}{\partial r}$ & $0$ & $0$\\
\end{tabular}
\right),
\end{equation}
\begin{equation}
\Delta\Omega\doteq \frac{1}{\mu_{0}}\theta(a-r)\left(
\begin{tabular}{ccc}
$ 0  $ & $~~0~~$ & $~~0~~$\\
$\frac{\partial}{\partial r}$ & $~~0~~$ & $~~0~~$\\
$\frac{1}{r}\frac{\partial}{\partial\phi}$ & $~~0~~$ & $~~0~~$\\
\end{tabular}
\right),
\end{equation}
and $\theta(x)$ is the unit step function with a value of one (zero) when $x>0$ ($x<0$). The operator $\Omega_{0}$ is the corresponding frequency operator for materials with isotropic permeability $\mu(r)$ and permittivity $\epsilon(r)$, where $\mu(r)=\mu'_{r}$ and $\epsilon(r)=\epsilon_{d}$ for $r<a$ and $\mu(r)=\epsilon(r)=1$ for $r>a$. For small $\mu_{k}\ll\mu_{r}$, we can consider the second term involving $\Delta\Omega$ as a first order perturbation to $\Omega_{0}$. By doing so, the effect of the static magnetic field is again seen as the two-step process shown in Fig. 1.

The frequency shift of each state in the first-order perturbation theory is $\Delta \omega_{nm}=\frac{\mu_{k}}{\mu_{k}^{2}-\mu_{r}^{2}}\langle\psi_{0,n,m}|\Delta\Omega|\psi_{0,n,m}\rangle$ (where $|\psi_{0,n,m}\rangle$ is the unperturbed state ket) and can be simplified as (see supplementary materials for derivations)
\begin{equation}
\Delta \omega_{nm}=2\pi c^{2} \cdot \frac{j_{z}|_{r=a}}{U} \cdot \frac{\mu_{k}}{\mu_{k}^{2}-\mu_{r}^{2}},\label{AM}
\end{equation}
where $j_{z}|_{r=a}=\frac{\epsilon_{0}m}{2\omega'_{n}}|E_{z}(k'_{n}a)|^{2}$ is the ``angular momentum density'' evaluated at the surface of the cylinder, $\omega'_{n}$ is the frequency of the unperturbed state, $k'_{n}=\omega'_{n}/c$, and $U=\frac{1}{2}\int[\epsilon_{0}\epsilon(r)|E|^{2}+\mu_{0}\mu(r)|H|^2] dA$. Here, the ``angular momentum density'' at the boundary is defined as $\vec{j}=\frac{\epsilon_{0}}{2\omega i} \vec{r}\times[\vec{E}^{*}\times(\nabla\times\vec{E})]$ \cite{LightAM}. In the non-dispersive approximation, $U$ can be considered as the total electromagnetic energy integrated on the $xy$-plane (see supplementary materials for a discussion on the region of integration). Since we consider frequencies far above $\omega_{h}$, the third factor in Eq. (\ref{AM}), $\mu_{k}/(\mu_{k}^{2}-\mu_{r}^{2})$, is positive and the sign of frequency shift is, therefore, the same as the sign of $m$. We thus conclude that the frequency shift is proportional to the ``angular momentum density'' at the cylinder surface per photon.

In comparison, the Zeeman effect on electronic states in atoms has a similar formula for the energy shift: $\Delta E=\mu_{B} g  J_{z} B/\hbar$, where $J_{z}$ is the total projected angular momentum in the direction of the static magnetic field ($B$), $\mu_{B}$ is the Bohr magneton, $\hbar=h/2\pi$ is the reduced Plank's constant, and $g$ is a dimensionless constant which depends on the type of atoms. Such an energy shift is proportional to the magnetic moments associated with the angular momentum of electrons. Here in Eq. (\ref{AM}), we also have the proportionality between frequency shift and angular momentum of light. With such proportionality, we thus call our work the Zeeman splitting of photonic angular momentum states in gyromagnetic cylinder, which is the most important result in this Letter. It should be noted that Eq. (\ref{AM}) are not limited to microwaves. One could easily extend this new theory to terahertz or optical frequencies in other systems, such as plasma systems where the roles of electric and magnetic fields are switched.

To verify our analytical formula [Eq. (\ref{AM})], we evaluate and compare it with the exact results in Fig. 2(c). Using the fact that the radiation field outside the gyromagnetic cylinder should not contribute significantly to the frequency shift, we get a closed-form expression for Eq. (\ref{AM}) (see supplementary materials for derivations):
\begin{equation}
\Delta\omega_{nm}\approx -\frac{\mu_{k}}{\mu_{r}}\frac{m\omega'_{n}}{k_{n}^{'2}a^{2}}\frac{J_{m}^{2}(k'_{n}a)}{J_{m}^{2}(k'_{n}a)-J_{m-1}(k'_{n}a)J_{m+1}(k'_{n}a)}. \label{ClosedForm}
\end{equation}
As shown in Fig. 2(c), we have a very good agreement between the frequencies evaluated from the closed-form solution [Eq. (\ref{ClosedForm})] and the exact results given by Eq. ({\ref{disp}}). Discrepancies exist only in the first few low-order resonances, which may be due to the first-order perturbation, the neglected radiation field, and the dispersive property of gyromagnetic materials. This proves the validity of Eq. (\ref{AM}).
\begin{figure}[htbp]
\centering
\includegraphics[width=3in]{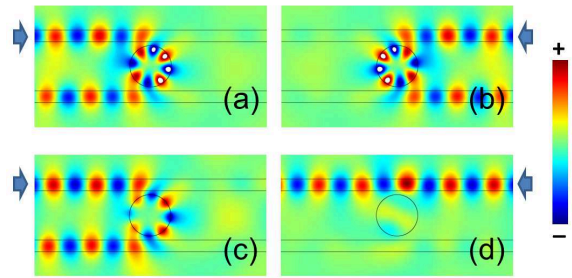}
\caption {\small (color online) Proposed experimental measurement for the Zeeman-like splitting of photonic angular momentum states. The cylinder in Fig. 1 is placed between two straight waveguides with a refractive index $2.2$ and width $5.6$ mm. The gap between the cylinder and each waveguide is $1.8$ mm. Arrows indicate the sources of the incoming waves. (a) and (b) Static magnetic field is absent. Waves launched from the left (a) and right (b) can also couple to the other waveguide. (c) and (d) Static magnetic field is present. Wave launched from the left (c) can couple to the other waveguide but that from the right (d) cannot.}
\end{figure}

With a non-degenerate angular momentum state that allows a rotation of electromagnetic field in only one direction, the Zeeman-like effect can lead to very unusual mode decoupling properties that could be useful for new photonic applications. For example, we consider the setup shown in Fig. 3. In the usual case, if we place the resonant cylinder between two waveguides, it can switch the light path from one waveguide to another waveguide through evanescent coupling. As shown in Fig. 3(a) and (b) for the cases without the static magnetic field, the device is a two-way coupler because the resonator supports both degenerate modes which couple the guided waves no matter which direction the source waves come from [see Fig. 3(a) for left-incident wave and Fig. 3(b) for right-incident waves]. However, by using the non-degenerate state $(1,-4)$, we find that the left-incident wave can couple to the other waveguide [Fig. 3(c)] but the right-incident wave fails to interact with the cylinder and pass straightly through the top waveguide [Fig. 3(d)]. The one-way transport properties can be understood by the mode coupling [Fig. 3(c)] and mode decoupling [Fig. 3(d)] between the incident guided wave and the single-mode state. The unusual properties shown in Fig. 3 can be used for experimental demonstration of the Zeeman-like splitting. Recently, the nonreciprocal transport properties of light in the presence of static magnetic field has been widely studied \cite{haldane,haldane2,zheng,ExpOneWay,alexander,eli,xianyu,shiyang}. The phenomena here may also be particularly useful for designing special photonic devices such as one-way coupled waveguides, diode for photonic circuit, or some add-drop filter systems \cite{zheng2,zongfu,kono,fan} with a much smaller region of applied static magnetic field (an area of one cylinder).

\begin{figure}[htbp]
\centering
\includegraphics[width=3.2in]{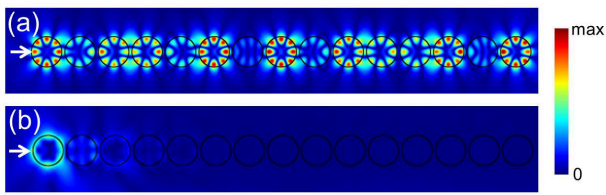}
\caption {\small (color online) Decoupling among photonic angular momentum states in an array of closely-packed identical gyromagnetic cylinders. The white arrows indicate the location of the point source.
(a) Strong coupling when static magnetic field is absent.
(b) Weak coupling when static magnetic field is present. }
\end{figure}

Another consequence of the Zeeman-like splitting is a localization due to the decoupling among closely spaced cylinders. As shown in Fig. 4, we can alter the interaction among an array of closely-packed resonators. In the usual case without the static magnetic field [Fig. 4(a)], the electric field can transfer through the array due to the strong coupling between cylinders. It is because each cylinder supports rotating states in opposite directions. When a clockwise rotational state is excited in one cylinder, the adjacent cylinder can attain the energy by its state in anti-clockwise rotation. The process keeps on at other cylinders and a guiding of wave is possible [shown as standing waves along the chain in Fig. 4(a)]. Such a guiding effect of coupled waves has been discussed in details before \cite{ResGuiding}. However, when we introduce the static magnetic field, only one rotational state is allowed in each cylinder at one frequency. When a clockwise rotational state is excited in one cylinder, the adjacent cylinder cannot attain the energy by its state in anti-clockwise rotation because this state is not allowed at the same frequency. This forces the electromagnetic energy almost localized in the first cylinder next to the source even when we tune the excitation frequency to the resonant frequency of these identical cylinders [Fig. 4(b)]. Such a great suppression of the interaction between resonators may be used for the design of slow-light devices and some applications where a high density of independent resonators is needed.

In conclusion, we showed analytically that the frequency splitting of the angular momentum states in a circular gyromagnetic cylinder under the presence of a static magnetic field is proportional to the angular momentum density of light at the surface of the gyromagnetic cylinder. These photonic angular momentum states give some unusual decoupling properties, which could be useful for designing novel photonic devices. Our study will benefit to the understanding of the angular momentum of light.

The authors are grateful to the financial support by NSF and the startup funding from MIT. J. Wang is supported by China Scholarship Council and Southeast University. We thank Anshuman Kumar, Dr. Zheng Wang, and Dr. Jack Ng for useful discussions.

\begin{figure}[htbp]
\includegraphics{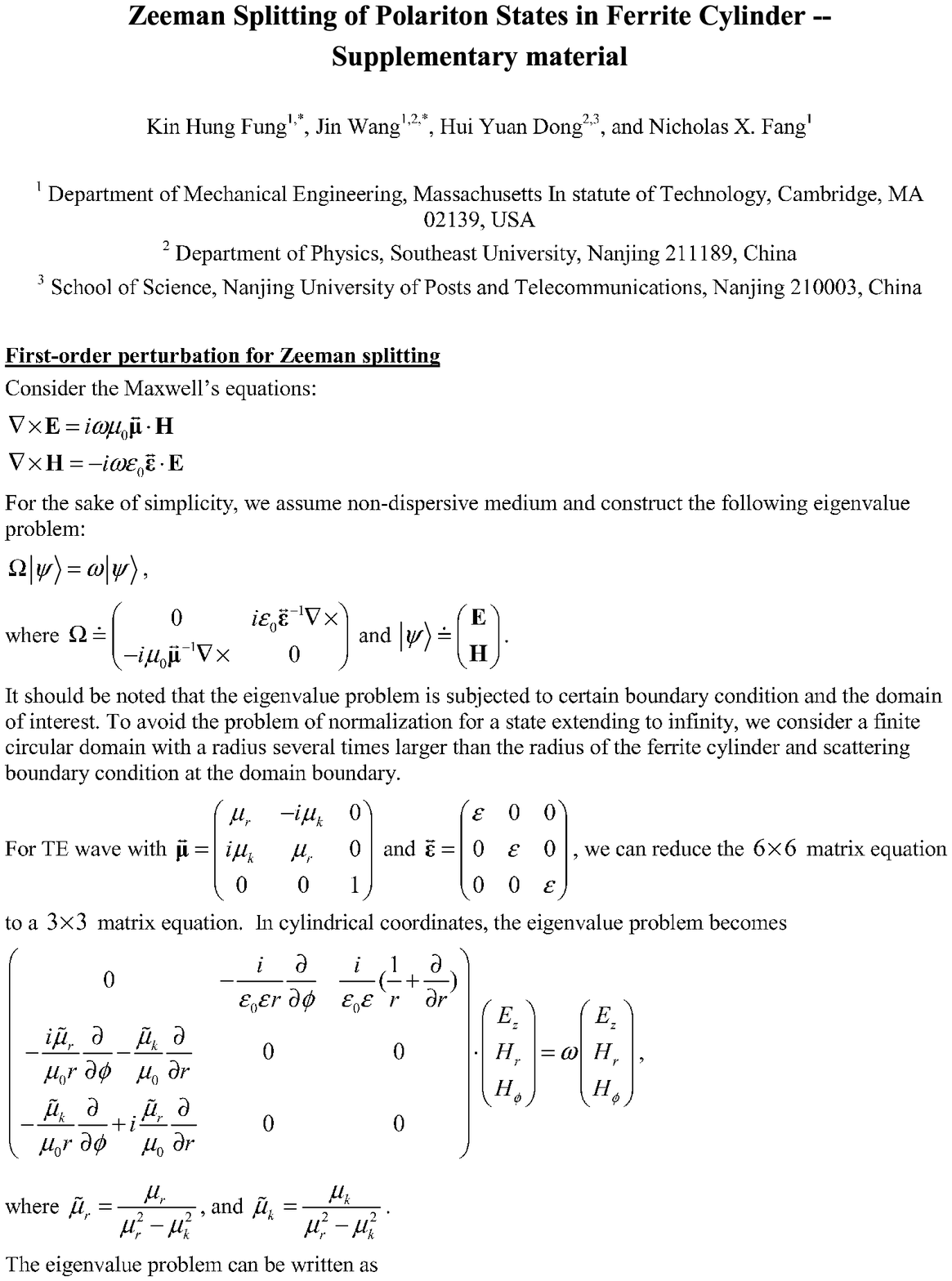}
\end{figure}

\begin{figure}[htbp]
\includegraphics{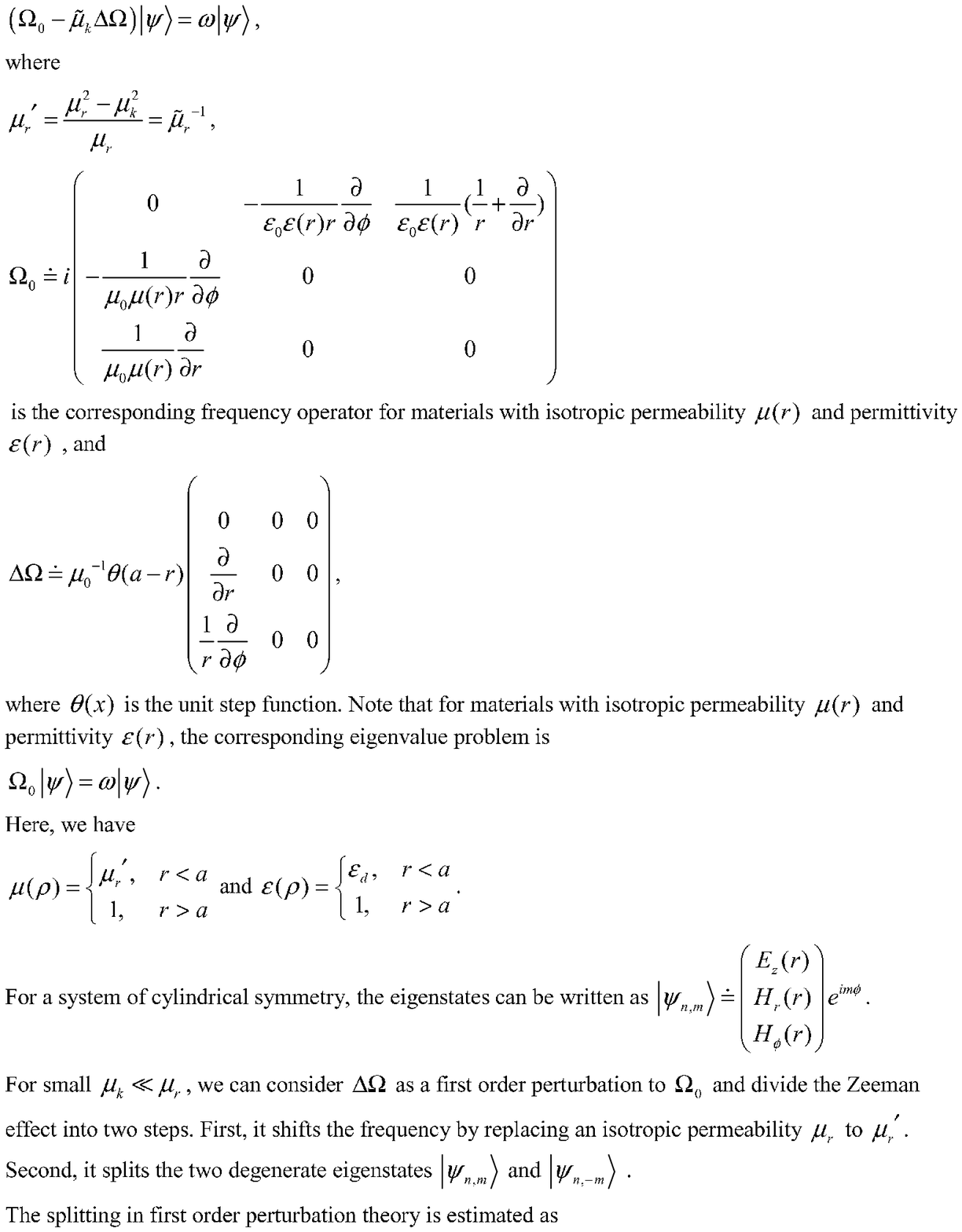}
\end{figure}

\begin{figure}[htbp]
\includegraphics{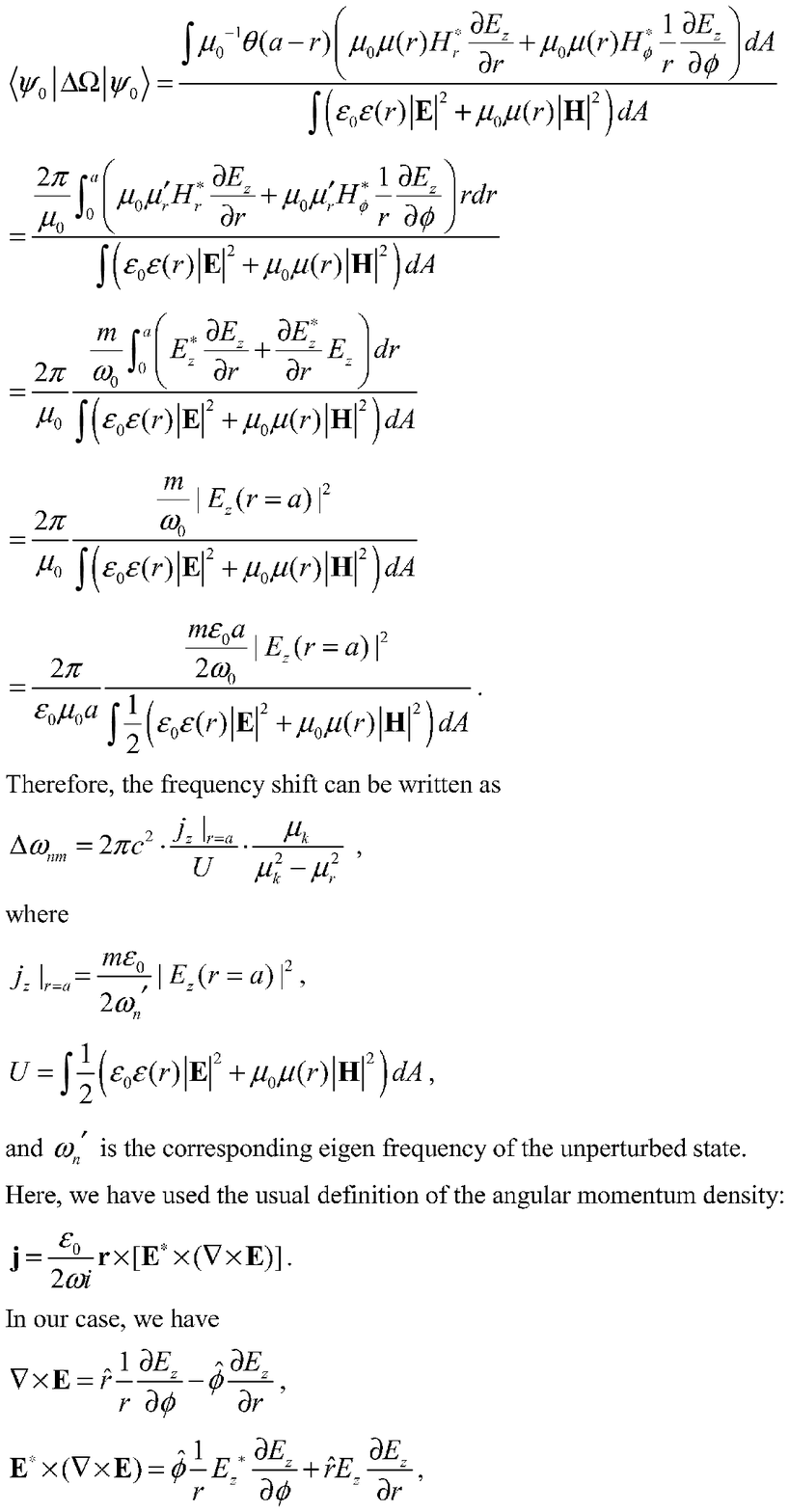}
\end{figure}

\begin{figure}[htbp]
\includegraphics{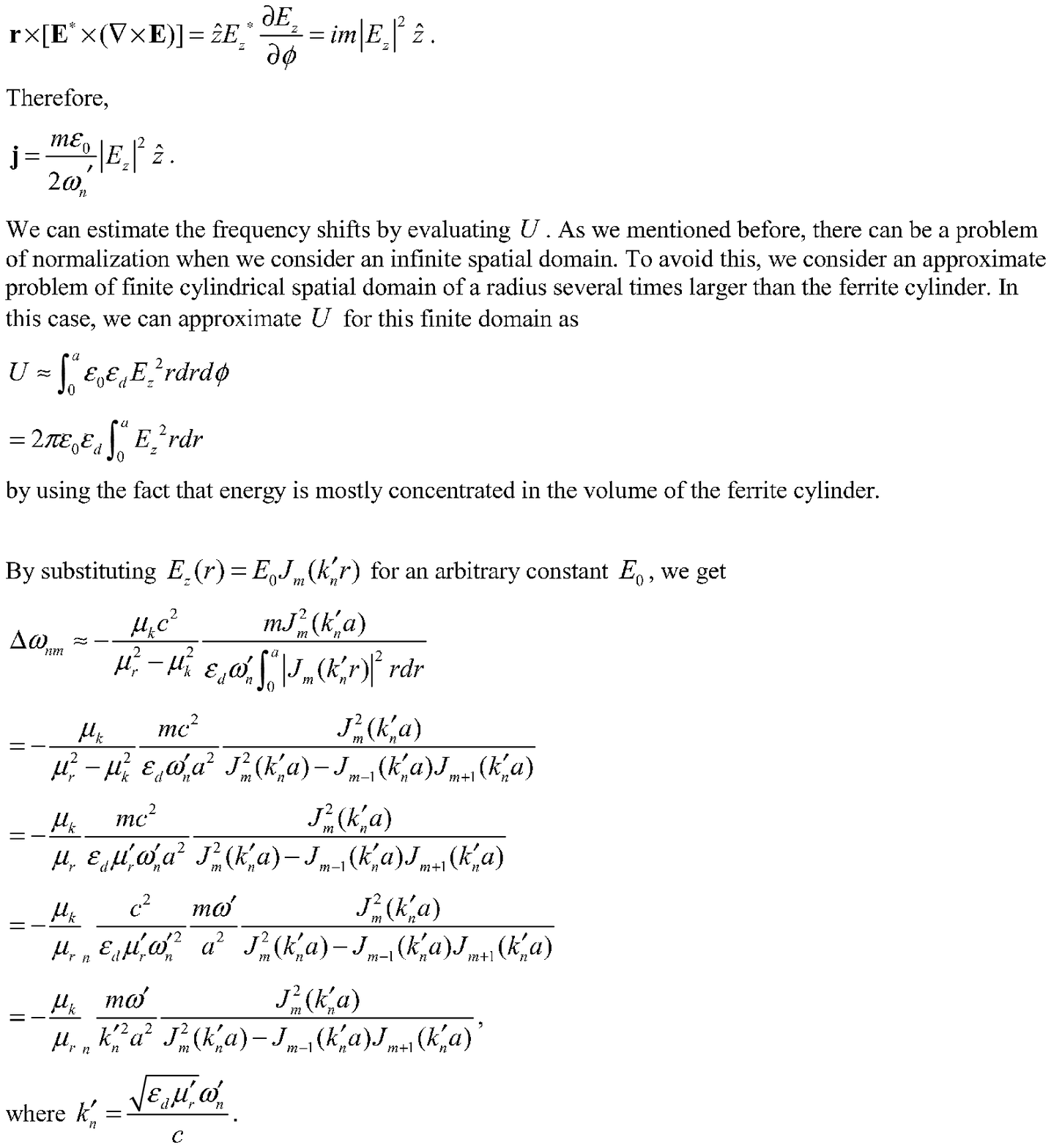}
\end{figure}

\end{document}